\documentclass[12pt,a4paper,final]{iopart}

\usepackage{iopams}  
\usepackage{graphicx}
\usepackage{cite}
\usepackage{epstopdf}
\usepackage{caption}
\usepackage[breaklinks=true,colorlinks=true,linkcolor=blue,urlcolor=blue,citecolor=blue]{hyperref}
\usepackage{booktabs}
\usepackage{float}
\usepackage{xcolor}

\begin{document}

\title[]{First principles investigation of topological phase in XMR material TmSb under hydrostatic pressure}

\author{Payal Wadhwa}
\address{Indian Institute of Technology Ropar, Rupnagar-140001, Punjab, India}

\author{Shailesh Kumar$^{1,2}$}
\address{$^1$School of Chemistry, Physics and Mechanical Engineering, Queensland University of Technology, Brisbane, Queensland 4000, Australia}
\address{$^2$Manufacturing Flagship, CSIRO, Lindfield West, New South Wales 2070, Australia}

\author{Alok Shukla}
\address{Indian Institute of Technology Bombay, Powai-400076, Mumbai, India}

\author[cor1]{Rakesh Kumar}
\address{Indian Institute of Technology Ropar, Rupnagar-140001, Punjab, India}
\eads{\mailto{rakesh@iitrpr.ac.in}}

\begin{abstract}
In this article, we report emergence of topological phase in XMR material TmSb under hydrostatic pressure using first principles calculations.\ We find that TmSb, a topologically trivial semimetal, undergoes a topological phase transition with band inversion at X point without breaking any symmetry under a hydrostatic pressure of 12 GPa.\ At 15 GPa, it again becomes topologically trivial with band inversion at $\Gamma$ as well as  X point.\ We find that the pressures corresponding to the topological phase transitions are far below the pressure corresponding to structural phase transition at 25.5 GPa.\ The reentrant behaviour of topological quantum phase with hydrostatic pressure would help in finding a correlation between topology and XMR effect through experiments.

\end{abstract}
\submitto{\JPCM}


\section{Introduction}

Recently, the family of rare earth monopnictides LnX (where, Ln is a rare earth element and X = As, Sb, Bi) have attracted great attention in the scientific community \cite{HULLIGER1977103,Petit_2016,tafti2016resistivity,sun2016large,PhysRevB.93.205152,neupane2016observation,ghimire2016magnetotransport,PhysRevB.95.014507,yu2017magnetoresistance,PhysRevB.96.235128,dey2018comparative,lou2017evidence,singha2017fermi} because of having remarkable signatures of extremely large magnetoresistance (XMR) \cite{PhysRevB.93.205152,ghimire2016magnetotransport,yu2017magnetoresistance,PhysRevB.96.235128,tafti2016resistivity,sun2016large} and superconductivity \cite{PhysRevB.95.014507,HULLIGER1977103}.\ Some LnX compounds like LaBi, CeBi, etc.\cite{lou2017evidence,singha2017fermi,PhysRevLett.120.086402} are observed to possess protected surface states in analogy with topological insulators \cite{RevModPhys.82.3045,RevModPhys.83.1057}.\ XMR is also observed in other non-trivial topological materials like WTe$_{2}$, Bi$_{2}$Te$_{3}$, TaAs, etc. \cite{chen2009experimental,ali2014large,PhysRevX.5.031023,PhysRevB.95.195113,fei2017edge}.\  Till now, the origin of XMR has been proposed through several mechanisms, for example perfect electron-hole compensation \cite{ali2014large,pletikosic2014electronic,guo2016charge,PhysRevB.97.235132,PhysRevLett.117.127204} and topological protection \cite{jiang2015signature,liang2015ultrahigh}.\ Further, the role of electron-hole compensation in explaining XMR is already understood by two-band model, in which compensation between electron and hole concentration leads to unsaturated quadratic behavior of MR, resulting in its dependence on the mobility of charge carriers \cite{ali2014large,pletikosic2014electronic,guo2016charge,PhysRevB.97.235132,PhysRevLett.117.127204}.\ However, the role of topology in explaining XMR is still not well established.\ In many recent reports, it is observed that topological protection suppresses backscattering at zero magnetic field and the application of external magnetic field breaks the topological protection, resulting in XMR \cite{jiang2015signature,liang2015ultrahigh}.\ But, there are many LnX compounds like LaSb, YSb, etc., which do not have topological protection still possess XMR \cite{PhysRevLett.117.127204,he2016distinct,xu2017origin}.\ Therefore, it is of great interest to explore a topologically trivial XMR material, which can become topologically non-trivial by increasing the strength of spin-orbit coupling (SOC), and may help in finding out a relation between topology and XMR effect.\ It is already known that the strength of SOC can be increased by applying pressure or strain or by chemical doping or alloying composition \cite{sato2011unexpected,brahlek2012topological,wu2013sudden,pal2014strain,yan2014experimental,sisakht2016strain,qi2017topological,mondal2019emergence}, amongst which external pressure is highly desirable tool as it does not affect the charge neutrality of the system.\\

Recent experimental reports have shown XMR in TmSb, which is topologically trivial semi-metal \cite{wang2018extremely}.\ In addition, TmSb is isostructural to LaSb/ LaAs, which are reported to undergo a topological phase transition under hydrostatic pressure \cite{guo2017theoretical,PhysRevB.98.220102}.\ It motivated us to investigate a topological phase in TmSb as a function of pressure.\ Further, LaSb and LaAs are found to undergo only a single topological quantum phase transition with pressure \cite{guo2017theoretical,PhysRevB.98.220102}.\ However, we observed two topological quantum phase transitions in TmSb under hydrostatic pressure, where it first becomes topologically non-trivial from trivial phase and again becomes topologically trivial from non-trivial phase within a pressure of 15 GPa.\ Therefore, the studies may help better in finding a correlation between topology and XMR effect through experiments.\\ 

\section{Computational details}

First principles calculations are performed within the framework of density functional theory (DFT) with projected augmented wave (PAW) formalism \cite{kresse1999ultrasoft} as implemented in VASP \cite{PhysRevB.54.11169}.\ Generalized gradient approximation (GGA) of Perdew-Burke-Ernzehrof (PBE) \cite{PhysRevLett.77.3865} as well as hybrid functionals (HSE06) \cite{heyd2003hybrid,brothers2008accurate} are used to include exchange-correlations.\ The cut-off energy of 300 eV is used for electronic band structure calculations.\ The sampling of the Brillouin zone (BZ) is done using k-mesh of size 11$\times$11$\times$11.\ The enthalpy of the system is calculated using GGA, under the hydrostatic pressure of 0-30 GPa, while band structure calculations are performed from 0-16 GPa.\ The system is simulated by variable cell relaxation under different applied pressures.\ The dynamical stability of the system under various pressures is confirmed by performing phonon dispersion calculations using PHONOPY code \cite{phonopy}.\ Band structure calculations are performed including the effect of spin-orbit coupling (SOC) after obtaining the crystal structure at different applied pressures.\ Z$_{2}$ topological invariant is calculated by Kane and Mele criterion \cite{PhysRevB.76.045302,PhysRevLett.98.106803}.\\    

\section{Results and discussions}

\begin{figure}[htb!]
 \centering
 \captionsetup{justification= raggedright,singlelinecheck = false}
 \includegraphics{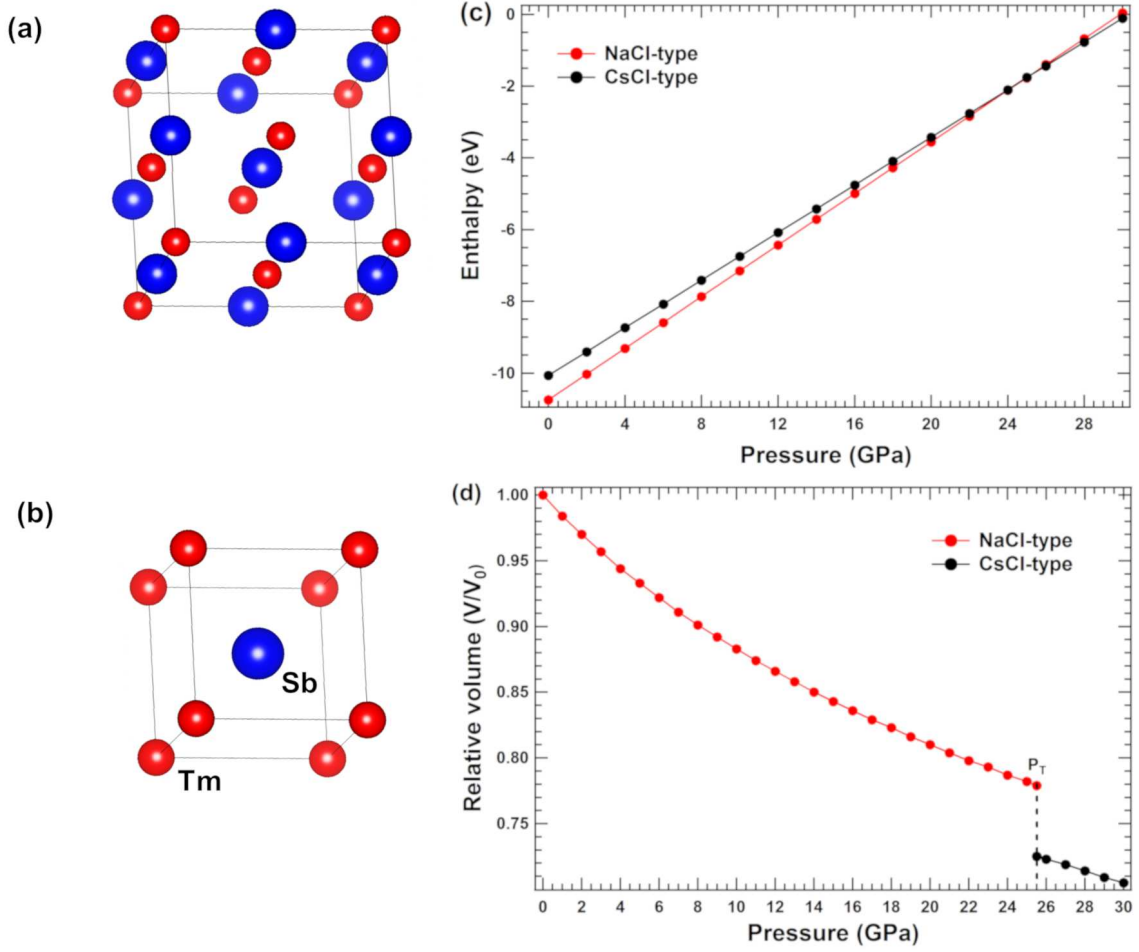}
 \caption{\label{figureone}(a) NaCl-type rocksalt crystal structure and (b) CsCl-type crystal structure of TmSb. Blue spheres represent Sb atom, while red spheres represent Tm atom. (c) Enthalpy as a function of pressure for NaCl and CsCl-type structures, (d) A change in the relative volume as a function of pressure in TmSb}
\end{figure}

To investigate a topological phase as a function of pressure in TmSb, we first checked stability of the crystal structure under pressures of 0-30 GPa.\ At ambient pressure, TmSb has NaCl-type rocksalt structure having space group Fm$\bar{3}$m (225) with Tm atom at the origin (0, 0, 0) and Sb atom at (0.5, 0.5, 0.5) (figure \ref{figureone}), which transforms into CsCl-type structure at higher pressure \cite{gupta2012structural}.\ The optimized lattice constant for TmSb at ambient pressure is found to be 6.131 \AA, which is in agreement with the previous reports (table \ref{zero}).\\ 
    
\begin{table}[H]
\centering
\caption{Comparison of lattice constant with the other reports}
\label{zero}
\begin{tabular}{@{}cccc@{}}
\toprule
\textbf{Material} & \textbf{PBE} & \textbf{LSDA+U} & \textbf{Expt.} \\ \midrule
TmSb              & 6.131        & 6.055 \cite{gupta2012structural}          & 6.105 \cite{ABDUSALYAMOVA1994107}        \\ \bottomrule
\end{tabular}
\end{table}

In order to ensure the structural phase transition (SPT), we calculated the enthalpies of both the crystal structures from pressures of 0-30 GPa [figure \ref{figureone}(c)].\ At a given pressure, the structure having lower enthalpy, H = E+PV (where E = total energy, P = external pressure, and V = volume of the unit cell) would have more stability.\ At ambient pressure, the enthalpy of NaCl-type structure of TmSb is lower than that of CsCl-type structure [figure \ref{figureone}(c)], indicating that former structure is more stable.\ On increasing pressure, the enthalpies of both the structures increases gradually and a crossover is observed at 25.5 GPa, which is called as transition pressure (P$_{T}$).\ It indicates that CsCl-type structure becomes more stable above 25.5 GPa, which is in close agreement with the earlier report \cite{gupta2012structural}.\  Further, the change in relative volume of TmSb is plotted as a function of hydrostatic pressure (Figure \ref{figureone}(d)).\ The abrupt change in volume at pressure P$_{T}$ = 25.5 GPa, indicates a first-order phase transition, which correspond to a change in the crystal symmetry \cite{gupta2012structural}.\ Moreover, the dynamical stability is confirmed by plotting its phonon dispersion spectrum under different applied pressures (supplementary material).\\

\begin{figure}[htb!]
 \includegraphics{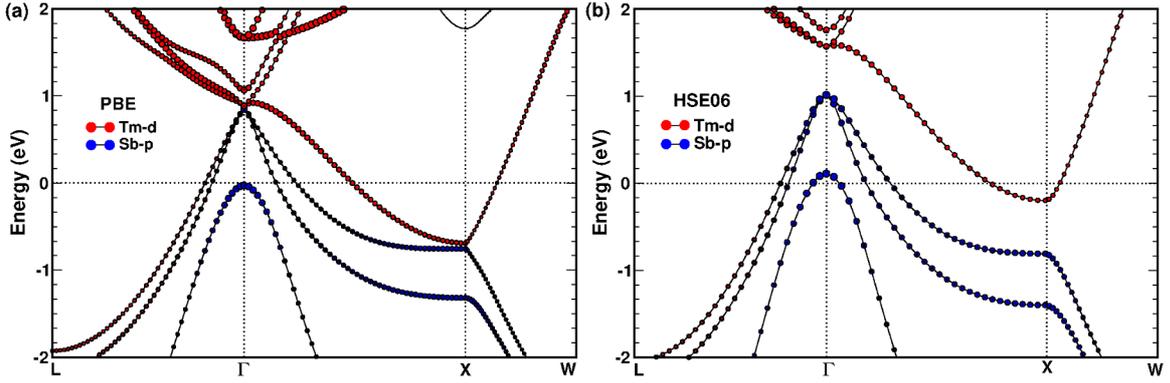}
 \captionsetup{justification= raggedright,singlelinecheck = false}
 \caption{\label{figuretwo}Band structures of TmSb including SOC effect using (a) PBE and (b) HSE06 functionals.\ Blue spheres show the contribution of \textit{p}-orbitals of Sb atom and red spheres show the contribution of \textit{d}-orbitals of Tm atom to the bands near the Fermi level.}
\end{figure}

For the investigation of topological phase in TmSb under hydrostatic pressure, first we calculated its band structure at ambient pressure using GGA and HSE06 functionals including the SOC effect.\ Band structure is plotted along the high symmetry points in the BZ, i.e., L to $\Gamma$, $\Gamma$ to X, and X to W, where we have excluded other high symmetry time-reversal variant points of K (0.375, 0.75, 0.375) and U (0.25, 0.625, 0.625), since we are interested only in topological phase transitions.\ Band structure plots for TmSb with SOC using GGA and HSE06 functionals are shown in figure \ref{figuretwo}.\\

It can be seen from the figure \ref{figuretwo} that valence band and conduction band near the Fermi level mainly have contributions from \textit{p}-orbitals of Sb (shown by blue spheres) and \textit{d}-orbitals of Tm (shown by red spheres) atoms, respectively, and a finite overlap is observed between the two using both GGA and HSE06 functionals.\ From its corresponding DOS plot (supplementary material), it is found to be semi-metallic, which is in agreement with the previous experimental reports \cite{wang2018extremely}.\ Further, no band inversion is found in their band structures calculated using both PBE and HSE06 functionals, indicating that it is topologically trivial at ambient pressure \cite{wang2018extremely}.\ It is to be noted that valence band maxima and conduction band minima almost touch at $\Gamma$ as well as X point in the band structures calculated using PBE functionals, while there is a finite gap between the bands at both $\Gamma$ as well as X point using HSE06 functionals.\ Since, an increase in pressure may lead to unphysical prediction of band inversion in case of band structures calculated using PBE functionals \cite{crowley2015accurate}, therefore we have used HSE06 functionals for our subsequent band structures calculations.\\ 

\begin{figure}[htb!]
 \centering
 \includegraphics{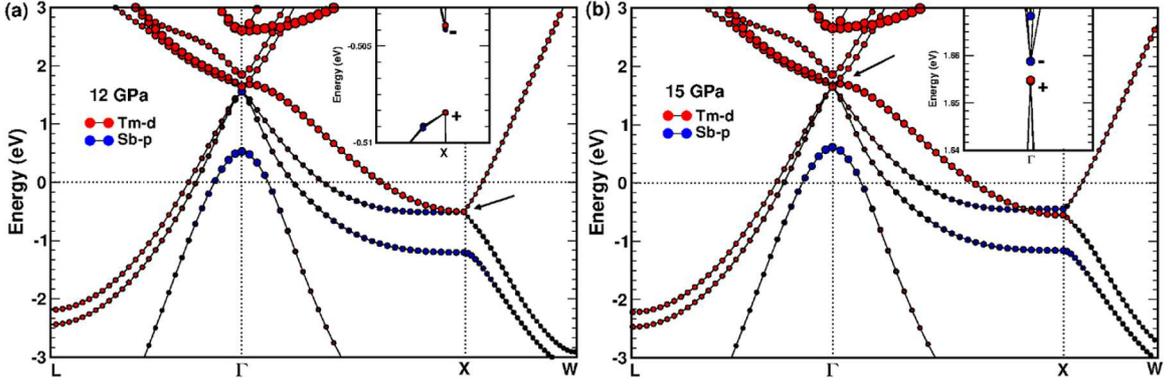}
 \caption{\label{figurethree}Band structures of TmSb using HSE06 functionals including SOC effect at pressure (a) 12 GPa and (b) 15 GPa.\ Blue spheres show the contribution of \textit{p}-orbitals of Sb atom and red spheres show the contribution of \textit{d}-orbitals of Tm atom near the Fermi level.}
\end{figure}

In order to explore topological phase in TmSb under pressure, we calculated its band structures from 0 to 16 GPa using HSE06 functionals including the SOC effect.\ To check non-trivial topological phase as a function of hydrostatic pressures, band inversion is checked at all the time-reversal invariant momenta (TRIM) points.\ Band inversion is found to be absent from 0 to 11 GPa. However, on increasing the pressure from 12-14 GPa, single band inversion is found at X point; but from 15 GPa onwards, two band inversions are observed at  $\Gamma$ as well as X point.\ It is to be noted that both of the pressures corresponding to topological  quantum phase transitions are far below the pressure of SPT at 25.5 GPa (figure \ref{figureone}).\ Band structure plots for TmSb at 12 GPa and 15 GPa are shown in figure \ref{figurethree}.\\ 

It is observed that at ambient pressure, the valence band at X point mainly has a contribution from the \textit{p}-orbitals of Sb atom, while the conduction band mainly has a contribution from the \textit{d}-orbitals of Tm atom.\ However, at X point a small contribution from \textit{d}-orbitals of Tm atom and \textit{p}-orbitals of Sb atom arise in the valence and conduction band, respectively, at 12 GPa, indicating a band inversion and the system becomes topologically non-trivial.\ At 15 GPa, two band inversions are observed at $\Gamma$ as well as X point (figure \ref{figurethree}). Even number of band inversions may correspond to either weak topological phase or topologically trivial phase \cite{PhysRevB.76.045302,PhysRevLett.98.106803}.\\ 

In order to further ensure the topological phase of TmSb under hydrostatic pressure, we calculated their Z$_{2}$ topological invariants as suggested by Kane and Mele \cite{PhysRevB.76.045302,PhysRevLett.98.106803}.\ For an inversion symmetric system in three dimensions, there exists four Z$_{2}$-invariants, i.e., ($\nu_{0}$: $\nu_{1}$, $\nu_{2}$, $\nu_{3}$), and the value of $\nu_{0}$ can be determined by the equation\

\begin{equation}
(-1)^{\nu_{0}}=\prod_{i}\delta _{i}
\end{equation}
   
where $\prod_{i}\delta _{i}$ denotes the product of the parities of all the filled bands at all TRIM points.\ In the three-dimensional BZ, there exists eight TRIM points.\ The value of first topological index $\nu_{0}$ = 1 corresponds to a strong topological phase and $\nu_{0}$ = 0 corresponds to either a weak topological phase or trivial insulator, which can be determined by the value of other three indices ($\nu_{1}$, $\nu_{2}$, $\nu_{3}$).\  The detailed parities of all the filled bands of TmSb at ambient pressure, 12 GPa, and 15 GPa are shown in table \ref{one}, \ref{two}, and \ref{three}, respectively.\\
 
\begin{table}[H]
\centering
\caption{Parities of all the filled bands at all the TRIM points in BZ at ambient pressure.}
\label{one}
\begin{tabular}{@{}cllllclllc@{}}
\toprule
\multicolumn{1}{l}{\textbf{Band No.}} & \textit{\textbf{L}} & \textit{\textbf{L}} & \textit{\textbf{L}} & \textit{\textbf{L}} & \multicolumn{1}{l}{\textit{\textbf{$\Gamma$}}} & \textit{\textbf{X}} & \textit{\textbf{X}} & \textit{\textbf{X}} & \multicolumn{1}{l}{\textbf{Total}} \\ \midrule
1                                     & -                   & -                   & -                   & -                   & -                                       & -                   & -                   & -                   & +                                  \\
3                                     & -                   & -                   & -                   & -                   & -                                       & -                   & -                   & -                   & +                                  \\
5                                     & -                   & -                   & -                   & -                   & -                                       & -                   & -                   & -                   & +                                  \\
7                                     & -                   & -                   & -                   & -                   & +                                       & +                   & +                   & +                   & +                                  \\
9                                     & +                   & +                   & +                   & +                   & -                                       & -                   & -                   & -                   & +                                  \\
11                                    & +                   & +                   & +                   & +                   & -                                       & -                   & -                   & -                   & +                                  \\
13                                    & +                   & +                   & +                   & +                   & -                                       & -                   & -                   & -                   & +                                  \\
Total                                 & +                   & +                   & +                   & +                   & +                                       & +                   & +                   & +                   & (+)                                  \\ \bottomrule
\end{tabular}
\end{table}

\begin{table}[H]
\centering
\caption{Parities of all the filled bands at all the TRIM points in BZ at 12 GPa.}
\label{two}
\begin{tabular}{@{}cllllclllc@{}}
\toprule
\multicolumn{1}{l}{\textbf{Band No.}} & \textit{\textbf{L}} & \textit{\textbf{L}} & \textit{\textbf{L}} & \textit{\textbf{L}} & \multicolumn{1}{l}{\textit{\textbf{$\Gamma$}}} & \textit{\textbf{X}} & \textit{\textbf{X}} & \textit{\textbf{X}} & \multicolumn{1}{l}{\textbf{Total}} \\ \midrule
1                                     & -                   & -                   & -                   & -                   & -                                       & -                   & -                   & -                   & +                                  \\
3                                     & -                   & -                   & -                   & -                   & -                                       & -                   & -                   & -                   & +                                  \\
5                                     & -                   & -                   & -                   & -                   & -                                       & -                   & -                   & -                   & +                                  \\
7                                     & -                   & -                   & -                   & -                   & +                                       & +                   & +                   & +                   & +                                  \\
9                                     & +                   & +                   & +                   & +                   & -                                       & -                   & -                   & -                   & +                                  \\
11                                    & +                   & +                   & +                   & +                   & -                                       & -                   & -                   & -                   & +                                  \\
13                                    & +                   & +                   & +                   & +                   & -                                       & +                   & +                   & +                   & -                                  \\
Total                                 & +                   & +                   & +                   & +                   & +                                       & -                   & -                   & -                   & (-)                                \\ \bottomrule
\end{tabular}
\end{table}

\begin{table}[H]
\centering
\caption{Parities of all the filled bands at all the TRIM points in BZ at 15 GPa.}
\label{three}
\begin{tabular}{@{}cllllclllc@{}}
\toprule
\multicolumn{1}{l}{\textbf{Band No.}} & \textit{\textbf{L}} & \textit{\textbf{L}} & \textit{\textbf{L}} & \textit{\textbf{L}} & \multicolumn{1}{l}{\textit{\textbf{$\Gamma$}}} & \textit{\textbf{X}} & \textit{\textbf{X}} & \textit{\textbf{X}} & \multicolumn{1}{l}{\textbf{Total}} \\ \midrule
1                                     & -                   & -                   & -                   & -                   & -                                       & -                   & -                   & -                   & +                                  \\
3                                     & -                   & -                   & -                   & -                   & -                                       & -                   & -                   & -                   & +                                  \\
5                                     & -                   & -                   & -                   & -                   & -                                       & -                   & -                   & -                   & +                                  \\
7                                     & -                   & -                   & -                   & -                   & +                                       & +                   & +                   & +                   & +                                  \\
9                                     & +                   & +                   & +                   & +                   & -                                       & -                   & -                   & -                   & +                                  \\
11                                    & +                   & +                   & +                   & +                   & -                                       & -                   & -                   & -                   & +                                  \\
13                                    & +                   & +                   & +                   & +                   & +                                       & +                   & +                   & +                   & +                                  \\
Total                                 & +                   & +                   & +                   & +                   & -                                       & -                   & -                   & -                   & (+)                                \\ \bottomrule
\end{tabular}
\end{table}

\begin{figure}[H]
 \centering
 \includegraphics{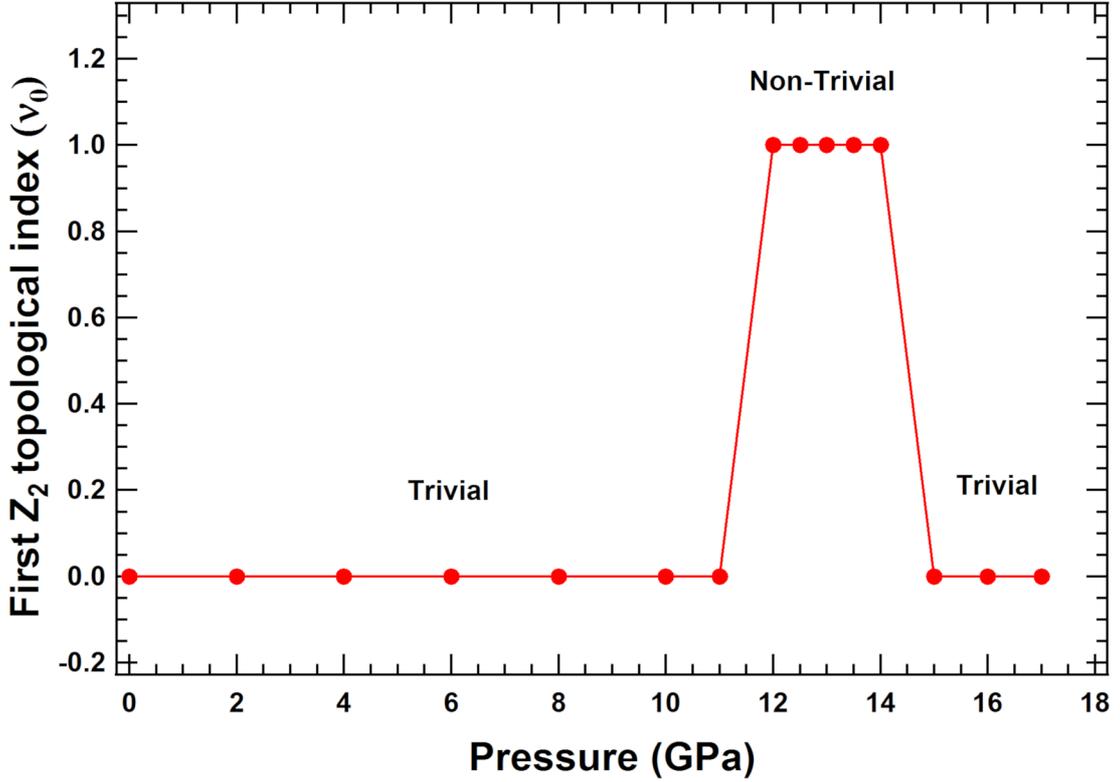}
 \caption{\label{figurefour}First Z$_{2}$ topological index ($\nu_{0}$) as a function of pressure in TmSb.}
\end{figure}

 Since, there is no band inversion at ambient pressure, and its first Z$_{2}$ index ($\nu_{0}$) calculated from equation (1) turns out to be zero, indicates that it is topologically trivial.\ At 12 GPa, band inversion at X point is observed and $\nu_{0}$ becomes 1, which indicates a topologically non-trivial phase.\ At 15 GPa, two band inversions are observed at $\Gamma$ as well as X points, turning $\nu_{0}$ to zero.\ The change in the value of $\nu_{0}$ from 1 to 0 shows that the system is converted either into a weak topological phase or topologically trivial phase.\ In order to ensure the topological phase at 15 GPa, we calculated the other three topological indices ($\nu_{1}$, $\nu_{2}$, $\nu_{3}$), which comes out to be (0, 0, 0) as the parity at four L and three X points are the same (table \ref{three}).\ It shows that the system has again become topologically trivial at 15 GPa.\ The value of the first Z$_{2}$ topological index as a function of pressure is plotted in figure \ref{figurefour}.\\
  
It is to be noted that with increase in pressure, an overlap between the valence and conduction band of TmSb increases, and would lead to an increase in the carrier concentration.\ Since XMR depends upon carrier concentration and mobility of the charge carriers, therefore evolution of mobility with pressure can only give the exact evolution of XMR as a function of pressure, which can only be explored in experiments.\ Consequently, the study of topological phase transitions in TmSb as a function of pressure provides a promising platform to investigate the role of topology in XMR effect. \\

\section{Summary}

On the basis of first principles calculations, it is summarised that XMR material TmSb undergoes a topological phase transition from trivial to non-trivial under a hydrostatic pressure of 12 GPa. On further increasing the pressure, it again becomes topologically trivial at 15 GPa.\ The studies indicate a reentrant topological quantum phase transitions in TmSb as a function of hydrostatic pressure, which may help to understand the correlation between topology and XMR effect through experiments.\\ 

\section*{Acknowledgement} 

The authors would like to thank IIT Ropar for High performance computing facility.

\section*{References}




\end{document}